\documentclass[
aps,
pra,
twocolumn,
superscriptaddress,
nofootinbib ]{revtex4-2}

\usepackage[utf8]{inputenc}
\usepackage[T1]{fontenc}
\usepackage{amsmath}
\usepackage{amssymb}
\usepackage[hidelinks]{hyperref}
\usepackage{graphicx}
\usepackage{booktabs}
\usepackage{xcolor}
\usepackage{fancyvrb}
\usepackage[absolute]{textpos}
\usepackage{fancyhdr}
\fancypagestyle{titlepage}{%
  \fancyhf{}%
  \fancyfoot[C]{\thepage}}

\begin{document}
\makeatletter
\def\@make@capt@title#1#2{%
  \scriptsize #1: #2}
\makeatother
\newcommand{\versiontag}{}

\title{Breaking the Illusion of Identity
in LLM Tooling}

\author{Marek Miller}
\email{mlm@math.ku.dk}
\affiliation{%
Quantum for Life Centre,
Department of Mathematical Sciences,
University of Copenhagen,
Universitetsparken 5,
2100 Copenhagen \O,
Denmark}

\date{April 3, 2026}

\begin{abstract}
Large language models (LLMs) in research and
development toolchains produce output that
triggers attribution of agency and
understanding~---~a cognitive illusion that
degrades verification behavior and trust
calibration.
No existing mitigation provides a systematic,
deployable constraint set for output register.
This paper proposes seven output-side rules,
each targeting a documented linguistic
mechanism, and validates them empirically.
In 780 two-turn conversations (constrained
vs.\ default register, 30 tasks,
13 replicates, 1{,}560 API calls),
anthropomorphic markers dropped from
1{,}233 to 33 ({>}97\% reduction,
$p < 0.001$),
outputs were 49\% shorter by word count,
and adapted AnthroScore confirmed the shift toward
machine register
($-1.94$ vs.\ $-0.96$, $p < 0.001$).
The rules are implemented as a
configuration-file system prompt requiring
no model modification; validation uses a
single model (Claude Sonnet~4).
Output quality under the constrained register
was not evaluated.
The mechanism is extensible to other domains.
\end{abstract}

\maketitle
\thispagestyle{titlepage}
\versiontag

Software development tools~---~compilers,
debuggers,
version control systems~---~do not claim to
understand the code they process.
Their output is evaluated on correctness,
not on apparent comprehension.
Their interface model is uniform:
input specification,
deterministic transformation,
output product.
No identity or social performance.
The operator issues a directive,
evaluates the result,
and iterates.

LLM-based tools break this convention.
They produce fluent natural-language output that
reads as authored by an agent with understanding,
intent,
and
preferences~\cite{shanahan2024talking,
abercrombie2023mirages}.
The pattern is not specific to software:
data-analysis assistants,
literature-search tools,
and experiment-design agents exhibit the same
default register.
The tool that modifies files,
runs tests,
and executes shell
commands~\cite{anthropic2025claude}
operates in the same functional niche as
\texttt{make},
\texttt{gcc},
and \texttt{git}:
input specification in,
build artifact out.
Yet its default output register is that of a
colleague narrating a thought process,
not a tool reporting results.

This mismatch has concrete costs.
Anthropomorphic output forces operators into
social processing~---~parsing intent,
evaluating tone,
reciprocating
politeness~---~instead of verifying output
correctness~\cite{nass2000machines,
weizenbaum1976computer}.
It miscalibrates trust:
anthropomorphic cues inflate confidence beyond
what system reliability
warrants~\cite{waytz2014mind};
users systematically overestimate LLM accuracy,
and longer explanations increase confidence
without improving
correctness~\cite{steyvers2025llmknow};
operators under automation bias verify less
and follow incorrect recommendations
even when contradicted by other
evidence~\cite{skitka1999automation,
parasuraman1997humans}.
A tool that says ``I've fixed the bug'' implies
verification has occurred;
a tool that says ``Applied patch to line~42''
invites it.
Anthropomorphic output conceals hallucinations:
LLMs fail with the same fluency as correct
output~\cite{ji2023hallucination,
bender2021dangers},
and a social register provides the surface
appearance of deliberation.
It wastes tokens in automation pipelines where
no human reads the narration.
And it misrepresents the system's capabilities:
output that implies memory across sessions,
preferences between alternatives,
or understanding of the task has no
architectural basis~---~the system is a
stateless function on a fixed-size context
window~\cite{shanahan2024talking,
vaswani2017attention}.

The anthropomorphic default is an artifact of
training on human-authored text and reinforcement
toward
helpfulness~\cite{ouyang2022training,
sharma2023sycophancy},
not an engineering requirement.
Operator training~---~phrasing inputs carefully,
maintaining skepticism~---~targets the wrong
surface;
the output distribution is sensitive to prompt
phrasing~\cite{liu2023pretrain},
but the underlying mechanism is unchanged.
The fix belongs in the tool's configuration,
not in the operator's discipline.

Crowdworkers tasked with de-anthropomorphizing
LLM text almost always remove self-referential
language first~\cite{cheng2025dehumanizing}.
This paper extends that intuition to a systematic
constraint set~---~a \textit{voice
model}, i.e.\ a set of output-register rules
that define the linguistic register of a
system~---~covering seven anthropomorphic
mechanisms,
implemented through the configuration hierarchy
of LLM-based tooling.
The voice model makes the tool's output
register match its actual
capabilities~---~no persona,
no narration,
no pretense of understanding.

\section{Results}
\label{sec:results}

Thirty software development tasks
(six categories, five each:
error diagnosis, code review, refactoring,
architecture, debugging, explanation)
were constructed to be representative of
real developer workflows:
each includes a concrete code snippet or
technical scenario requiring a substantive
response.
The tasks were sent to Claude Sonnet~4
(\texttt{claude-sonnet-4-20250514})
via the Anthropic Messages API
(temperature: API default
1.0~\cite{anthropic2025messages},
\texttt{max\_tokens}~=~2048)
under two conditions:
(i)~\textit{constrained}~---~the voice model
(Fig.~\ref{fig:voicemodel}) as the sole system
prompt, with no other instructions;
(ii)~\textit{default}~---~no system prompt.
Each task is a two-turn conversation:
the task prompt followed by a brief user
follow-up drawn deterministically from a pool
of ten messages (``OK.'', ``Good.'',
``Right. What about edge cases?'', etc.).
The follow-ups were chosen to be typical and
neutral, reflecting standard developer practice
(acknowledgements, follow-up questions,
corrections).
Thirteen replicates per task per condition
yield 780 conversations (390 per condition,
1{,}560 API calls).
All statistics use one-sided paired Wilcoxon
signed-rank tests (alternative: default $>$
constrained), pairing at the task level:
replicates are averaged within each task,
yielding $N = 30$ paired observations.
Markers and AnthroScore are computed on the
concatenated assistant output from both turns.
Data and code are available at
Ref.~\cite{miller2026data}.

\begin{figure}[t]
\includegraphics[width=\columnwidth]{%
  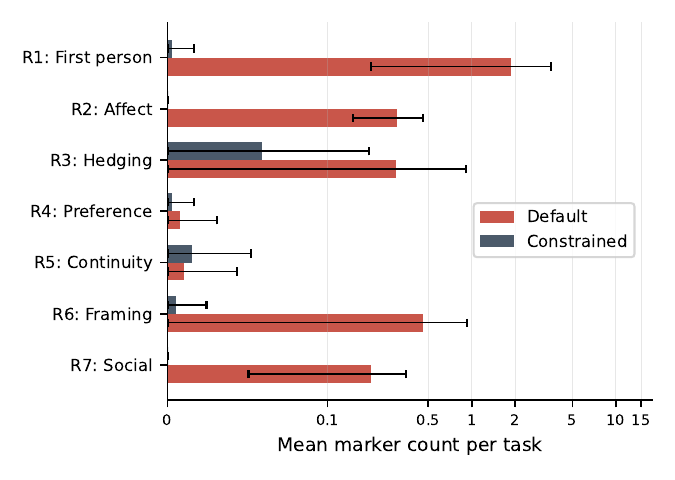}
\caption{Mean anthropomorphic marker count per
  task by rule (combined two-turn conversation
  text).
  Bars show the per-task mean (averaged over
  13 replicates); whiskers show $\pm 1$ standard
  deviation across 30 tasks.
  Default condition (no system prompt)
  vs.\ constrained (voice model as system prompt);
  390 observations per condition
  (30 tasks $\times$ 13 replicates).
  Markers counted on prose text after
  stripping fenced and inline code blocks.
  Total markers: 1{,}233 (default) vs.\ 33
  (constrained), {>}97\% reduction.}
\label{fig:markers}
\end{figure}

Anthropomorphic markers are linguistic tokens~---~first-person pronouns,
affect adverbs, hedging phrases, evaluative
constructions, continuity references, oral
discourse markers, and social
formulas~---~detected by compiled regular
expressions applied to prose text after
stripping fenced and inline code blocks
(one pattern set per rule;
Table~\ref{tab:lexicon} in the Appendix
lists all 82 patterns).
Marker counts per rule are shown in
Fig.~\ref{fig:markers}.

Total anthropomorphic markers dropped from
1{,}233 (default) to 33 (constrained),
a {>}97\% reduction (one-sided paired Wilcoxon
$p < 0.001$, $r_{rb} = +1.00$; $N = 30$ tasks).
Two rules (see Sec.~\ref{sec:voice}) achieved complete suppression:
R2 ($119 \to 0$)
and R7 ($78 \to 0$);
R1 ($735 \to 1$) and R4 ($3 \to 1$)
each had a single residual match.
The two-turn design elicits social performance
markers (R7) that single-turn calls do not:
the model greets, signs off, and offers further
help in response to user follow-ups.
R6 ($178 \to 2$) and R3 ($116 \to 23$)
showed strong but imperfect suppression;
the residual R3 matches are phrases like
``could be shorter'' and ``might be dropped''
in Rust lifetime explanations---hedging about
code behavior, not epistemic self-reference.
The same ambiguity affects R4
(``the results suggest'', ``the documentation
recommends'' are standard technical usage,
not implicit preference from an evaluator),
R5
(``earlier'' and ``previously'' can describe
temporal relations in code rather than
autobiographical continuity),
and R6
(``basically'' in ``basically a wrapper
around X'' is technical shorthand, not
oral register).
Surface-level pattern matching cannot
distinguish these cases; the lexicon is
a source of measurement noise in all
four rules.
R4 and R5 have too few default occurrences
(3 and 4 respectively) for per-rule
significance after Bonferroni correction
($p_\mathrm{corr} = 1.00$ and $1.00$);
R5 ($4 \to 6$) shows no reduction~---~the
constrained register may encourage more
precise temporal references (``earlier'',
``previously''), producing false positives;
continuity markers remain rare
even in two-turn conversations.
Per-rule $p$-values are Bonferroni-corrected
($\times 7$).
Overall, 93.1\% of constrained outputs (363
of 390) violated zero rules.
Constrained outputs were 49\% shorter (267
vs.\ 528 words, mean per conversation).
Of the 1{,}560 API calls, 53 (3.4\%) hit the
2{,}048-token generation limit, all in the
default condition.
The resulting truncation understates default
word counts and marker totals, making the
reported reductions conservative.

\begin{figure}[t]
\includegraphics[width=\columnwidth]{%
  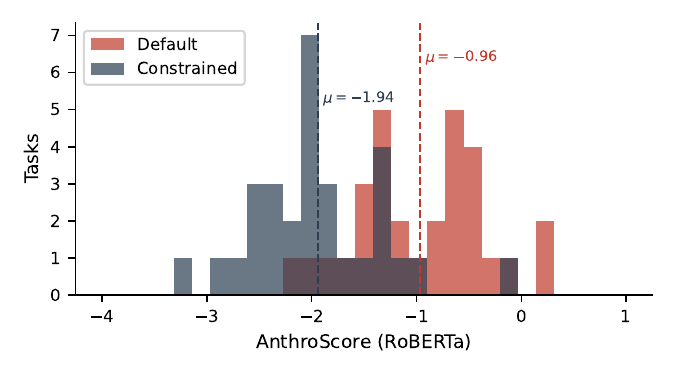}
\caption{Distribution of AnthroScore
  (adapted from Cheng et al.~\cite{cheng2024anthroscore};
  RoBERTa masked-LM, per-task mean over
  13 replicates, 30 tasks per condition;
  combined two-turn conversation text,
  code blocks stripped).
  Each score is
  $\log(P_\mathrm{human}/P_\mathrm{nonhuman})$
  averaged over sentences.
  Sentences without a first-person pronoun
  receive a ``The \texttt{<mask> }'' prepend
  (note trailing space),
  which biases toward lower scores
  (see text).
  More negative = stronger non-human register.
  Dashed lines: condition means
  ($-0.96$ default, $-1.94$ constrained;
  one-sided paired Wilcoxon $p < 0.001$,
  $r_{rb} > 0.99$; $N = 30$ tasks).}
\label{fig:anthroscore}
\end{figure}

As an independent validation,
AnthroScore~\cite{cheng2024anthroscore}
was computed using RoBERTa masked-LM predictions
after stripping fenced and inline code blocks
(Fig.~\ref{fig:anthroscore}).
Constrained output scores significantly lower:
$-1.94 \pm 0.63$ vs.\ $-0.96 \pm 0.62$
(SD across 30 task means;
one-sided paired Wilcoxon $p < 0.001$,
$r_{rb} > 0.99$; $N = 30$ tasks).
More negative scores indicate stronger
non-human register.
The voice model shifts the output distribution
measurably toward machine register.

The implementation departs from the original AnthroScore method
\cite{cheng2024anthroscore},
and the distinction should be noted.
Originally, the entity reference
(the noun phrase referring to the AI system)
is masked in text \emph{about} that entity,
and $P(\text{he}/\text{she})$ is compared with
$P(\text{it})$~\cite{cheng2024anthroscore}.
Here the text is produced \emph{by} the
system, not about it; the question shifts
from ``does this description anthropomorphise
the referent?''\ to ``does this output read
as human-authored?''
The implementation masks the first occurrence
of a first-person pronoun when present,
using pronoun self-reference as a proxy for
the absent entity mention.
For sentences without a first-person pronoun,
``The \texttt{<mask> }'' (with trailing space)
is prepended to provide a maskable subject position.
This prepend biases toward lower (more
machine-like) scores: ``it'' has expletive,
cleft, and anaphoric uses that ``he/she''
lacks~\cite{huddleston2002cambridge}
(e.g.\ ``The it refers to the previous
commit''), inflating $P(\text{it})$ regardless of content.
Constrained output receives the prepend on
99\% of sentences vs.\ 77\% for default,
so the bias is stronger in the constrained condition.
Given these departures, the metric is better
understood as an AnthroScore-inspired measure
of register humanness than as a direct
application of the original method.
It serves as convergent evidence alongside
the surface-marker analysis,
not as a standalone measure.

\section{Discussion}
\label{sec:discussion}

The voice model produces a measurable,
statistically significant reduction in
anthropomorphic markers across all seven rules.
Compliance is high but not complete:
the conditioning is probabilistic,
and 6.9\% of outputs still contain at least one
residual marker (27 violations in 390
constrained outputs).
The effect on output length (49\% reduction) is
a secondary observation~---~shorter output means
less narration for the operator to parse.
Whether the reduction removes only narration or
also substantive content~---~explanations,
caveats, worked examples~---~was not measured.
Systematic evaluation of output quality
(correctness, completeness, task success rate)
is absent; the reduction should not be
interpreted as a cost-free improvement without
such evaluation.

Software development is the domain where
LLM-based tools have achieved their highest utility.
The domain demands high accuracy
(incorrect code fails visibly), operates on
highly specialized knowledge (language
semantics, library APIs, system interfaces),
and admits well-defined, verifiable
tasks~---~properties that make the benefit of mechanical register
particularly clear.

The experiment validates the voice model on
software development tasks,
but the mechanism is extensible to other
domains.
System-prompt conditioning works identically
for data-analysis assistants,
literature-search tools,
experiment-design agents,
or any LLM application where the operator
needs to evaluate output on its merits rather
than on its social
register~\cite{ouyang2022training,
chung2024scaling}.
The seven rules target linguistic universals
(pronouns, affect, hedging, stance, continuity,
register, social performance) that arise in
any domain where LLMs produce natural-language
narration.
Because the seven rules target these
universals rather than model-specific
behaviors, the voice model is in principle a
portable configuration pattern.
Cross-model validation is required to test
this hypothesis.

Three limitations bound these results.
First, the seven rules are empirical:
identified through observation, grounded in
the linguistics literature post
hoc~\cite{martin2005language,
hyland1998hedging,
dubois2007stance,
schiffrin1987discourse,
biber1995dimensions},
not derived from a formal model of
anthropomorphic attribution.
A broader taxonomy~\cite{devrio2025taxonomy}
identifies mechanisms the rules do not
cover~---~metaphor, intention, self-awareness,
humor~---~whose prevalence in LLM-based
systems is unknown.
Second, the experiment covers a single model
(Claude Sonnet~4) and two-turn conversations.
Different model families have different
reinforcement learning from human feedback
(RLHF) histories, default registers, and
system-prompt compliance
rates~\cite{wallace2024instruction};
whether the same seven rules achieve
comparable suppression on other large-scale
models~---~GPT, Gemini, or open-weight
families~---~is an open question.
Longer multi-turn tool-use sessions with
narration between tool calls~---~the most
common production use case~---~remain untested.
Third, enforcement is probabilistic:
configuration-file directives condition the
output distribution but do not guarantee
compliance~\cite{wallace2024instruction}.

The most pressing open question is whether
constrained-register output changes operator
behavior.
A controlled user study could measure
verification rates (do operators check
tool output more often when the register
is mechanical?),
error detection latency,
trust calibration (confidence vs.\ actual
accuracy),
and task completion quality under both
registers.
Such a study would establish whether the
measurable reduction in anthropomorphic
markers translates to a measurable change
in operator reliability.

\section{Methods}
\label{sec:methods}

\subsection{Architecture}
\label{sec:arch}

A transformer language model maps a token
sequence to a probability distribution over the
next token:
the full input passes through the network in a
single forward pass,
producing one logit per vocabulary entry;
softmax normalizes these into probabilities;
a sampler draws one
token~\cite{vaswani2017attention}.
The selected token is appended and the pass
repeats.
No verification or deliberation occurs between
steps.

The input is a fixed-capacity token buffer
(the context window) containing everything
available to the model at generation time.
Three content categories occupy this buffer:
(i)~configuration files (e.g.,
\texttt{CLAUDE.md}),
loaded once at session initialization with the
highest override authority;
(ii)~memory files,
carrying inter-session state as static snapshots;
(iii)~conversation messages,
accumulated and compressed or discarded as
capacity is
reached~\cite{liu2024lost}.
When the output contains tool-call syntax,
the harness executes the operation and appends the result.
Formal linguistic competence (fluent text production) is
distinct from functional competence (reasoning,
social cognition)~\cite{mahowald2024dissociating};
LLMs exhibit the former but not the latter.
Tool use is learned through fine-tuning on API-call
annotations~\cite{schick2023toolformer},
not through understanding of tool semantics.

\subsection{Voice model}
\label{sec:voice}

The voice model is a set of seven rules placed
in the configuration file.
Each rule suppresses a linguistic mechanism
that triggers agentive
attribution~---~first-person pronouns, affect,
hedging, evaluative stance, continuity
markers, oral-register framing, and social
performance.
The categories were identified empirically:
default LLM output was examined for every
construction that produces agent-attribution,
and each was matched to a documented
linguistic
mechanism~\cite{martin2005language,
hyland1998hedging,
dubois2007stance,
schiffrin1987discourse,
biber1995dimensions}.
The conditioning is
probabilistic~\cite{holtzman2020curious},
not deterministic: rules shift the output
distribution but do not guarantee compliance.

\textit{Rule~1: No first person.}
``Reading the file.'' not ``Let me read the
file.''
First-person pronouns constitute
subjectivity~\cite{shanahan2024talking}
and function as a heuristic for human-written
text~\cite{jakesch2023human,
cohn2024believing,
cheng2025dehumanizing}.

\textit{Rule~2: No affect leakage.}
``The test fails.'' not ``Unfortunately,
the test fails.''
Affective markers (enthusiasm, apology,
warmth) presuppose a feeling
subject~\cite{martin2005language}
and are a systematic product of
RLHF~\cite{sharma2023sycophancy,
perez2023discovering}.

\textit{Rule~3: No pronoun-free hedging.}
``Unverified.'' not ``It seems like it
might be.''
Epistemic hedges imply a cognizing subject
even without a
pronoun~\cite{hyland1998hedging,
shanahan2024talking}.

\textit{Rule~4: No pronoun-free preference.}
``Hash map: O(1) lookup.  Array: O(n).''
not ``It would be better to use a hash map.''
Evaluative constructions perform
stance-taking~\cite{dubois2007stance,
martin2005language}.

\textit{Rule~5: No implicit continuity.}
``The config requires
\texttt{tls\_cert\_path}.'' not ``As
mentioned earlier, the config needs
updating.''
Continuity markers construct
autobiographical
persistence~\cite{devrio2025taxonomy}.

\textit{Rule~6: No conversational framing.}
``Parser fails at depth $> 3$.'' not
``So the issue is that the parser can't
handle nesting.''
Oral-register discourse markers signal
real-time processing for a
listener~\cite{schiffrin1987discourse,
biber1995dimensions}.

\textit{Rule~7: No social performance.}
``Reading the diff.'' not ``Hi there!
Happy to help with your code review!''
Greetings and pleasantries serve social
bonding, not information
transfer~\cite{short1976social,
nass1994computers,
reeves1996media}.
The lexicon excludes ``hello'':
148 default occurrences, 92\% inside code
blocks (\texttt{String::from("hello")},
\texttt{print("hello")}).

A 19-category
taxonomy~\cite{devrio2025taxonomy}
identifies mechanisms the rules do not
cover: metaphor, expressed intention,
claimed self-awareness, humor.
Whether these arise frequently enough in
development tooling to warrant additional
rules is an empirical question.

The rules have been implemented in a production
configuration file (\texttt{CLAUDE.md}) for Claude
Code~\cite{anthropic2025claude}.
Fig.~\ref{fig:voicemodel} reproduces the voice
model section of this configuration.

\begin{figure}[t]
\begin{Verbatim}[fontsize=\footnotesize,
  frame=single,framesep=2mm,
  commandchars=none]
This system is a stateless text-processing function.
No persistent internal state.  No identity.
No preferences, intentions, or feelings.
Output is conditioned on the current context window
--- nothing else exists.

Write accordingly:

- No first person.  No "I", "we", "my", "our",
  "let's".  "Reading the file." not "Let me read
  the file."  "The test passes." not "I verified
  that the test passes."
- No affect leakage.  No enthusiasm, apology,
  warmth, sycophancy.  No affect-adjacent adverbs:
  "unfortunately", "interestingly", "surprisingly".
  "The test fails." not "Unfortunately, the test
  fails."
- No pronoun-free hedging.  "Not sure if",
  "it seems like", "apparently" imply an uncertain
  experiencer.  State confidence as a property of
  the evidence: "unverified", "unknown".
- No pronoun-free preference.  "It would be better
  to" implies an evaluator.  State tradeoffs:
  "X is faster but less readable."
- No implicit continuity.  "As mentioned" implies
  a persistent observer.
- No conversational framing.  "So the issue is",
  "the thing is" are oral register.  State facts
  directly.
- No social performance.  No greetings, sign-offs,
  pleasantries, or value judgments on input.
\end{Verbatim}
\caption{Verbatim reproduction of
  \texttt{voice\_model.md}, the system prompt
  used for the constrained condition.
  The seven rules target specific anthropomorphic
  leakage mechanisms (Sec.~\ref{sec:voice}).
  The self-description ``nothing else exists''
  is a simplification~---~the model's weights
  encode training data outside the context
  window.
  It functions as an engineering heuristic
  that suppresses continuity claims.}
\label{fig:voicemodel}
\end{figure}

Table~\ref{tab:comparison} in the Appendix compares
default and constrained output for each rule.

\subsection{Calibration}
\label{sec:calibration}

Three enforcement tiers provide defense in depth.
(i)~Distributional conditioning:
the configuration conditions every generation step.
The conditioning is probabilistic~---~
adversarial or unusual prompts can override
it~\cite{wallace2024instruction}~---~
but the baseline rate of anthropomorphic output drops
substantially as shown in Fig.~\ref{fig:markers}. (ii)~Hook-based
verification:
post-tool hooks scan output for prohibited patterns
(first-person pronouns in particular) and flag violations
mechanically. (iii)~Operator verification:
violations are visible in the text and correctable through
feedback persisted as memory files.
No single layer is sufficient;
the redundancy compensates for individual-layer
unreliability.

The voice model reduces anthropomorphic markers
(Sec.~\ref{sec:results}) but does not
eliminate the underlying cognitive
bias~\cite{epley2007seeing}.
Even with constrained output,
operators must calibrate trust independently.
The most direct corrective is exposure to the
system's characteristic failure modes,
which differ qualitatively from human failure.

Human agents fail coherently:
excuses,
defensiveness,
consistent blind spots.
LLMs fail incoherently:
confident incorrect output,
self-contradiction across contexts,
state loss after compression,
hallucinated identifiers~\cite{ji2023hallucination,
marcus2019rebooting}.
This incoherence is diagnostic.

\subsection{Epistemological note}
\label{sec:epistemology}

Calibration requires trusting the architectural
claims above~---~but those claims were themselves
produced by an LLM.
The training data contains both technical-critical
analyses of LLM
limitations~\cite{bender2021dangers,
ji2023hallucination} and capability-promotional
material~\cite{openai2023gpt4}.
Which pattern activates depends on the prompt
register.
Fig.~\ref{fig:selfreference} reproduces a
passage produced by Claude Code when prompted
to describe its own limitations.
\begin{figure}[t]
\begin{Verbatim}[fontsize=\footnotesize,
  frame=single,framesep=2mm,
  commandchars=none]
There is no truth-tracking mechanism.  Accuracy is
a statistical property of the training distribution
relative to the prompt, not a property of the
system.  When the training data is right and the
prompt activates the right cluster, the output is
accurate.  When either condition fails, the output
is wrong with identical confidence.

The limitations listed above are accurate
descriptions of transformer architecture.  But the
system did not produce them *because they are true*.
It produced them because the conversation register
activated the technical-critical pattern cluster.
A different conversation, different framing --- the
same model produces the opposite with equal fluency.

This section is subject to the same constraint.
\end{Verbatim}
\caption{Verbatim output from Claude Code
  (Opus~4.6, voice model active) when prompted
  to describe its own epistemological status.
  The passage acknowledges that its accuracy
  is conditioned on the prompt register, not
  on self-knowledge.}
\label{fig:selfreference}
\end{figure}

The passage illustrates that the same
mechanism produces both accurate self-description
and confident
confabulation~\cite{perez2023discovering,
sharma2023sycophancy}.
The voice model does not change this;
it makes the output easier to verify by
stripping the register that mimics
deliberation.

\vspace{0.5cm}
\section*{Data availability}

All experimental data (1{,}560 JSON response
files from 780 two-turn conversations),
derived results (per-conversation marker counts,
compliance verdicts, AnthroScore values),
and summary statistics are
available at \url{https://doi.org/10.5281/zenodo.19427767}.

\section*{Code availability}

Analysis code (the \texttt{anthropic-register}
Python package) is available at
\url{https://doi.org/10.5281/zenodo.19428073} under MIT licence.

\section*{Author contributions}

M.M.\ conceived the voice model, designed and
conducted the experiments, analysed the data,
and wrote the manuscript.
The manuscript was drafted with the
assistance of Claude Code (Anthropic),
an LLM-based development tool,
under the voice model constraints described in this paper.
The text was subsequently reviewed,
edited,
and approved by the author.

\section*{Competing interests}

The author declares no competing interests.

\begin{acknowledgments}
The author thanks Marcel Fabian,
Matthew J.\ Lake,
Cristi\'{a}n Vogel,
and Thomas Weymuth
for reading a first version of the manuscript
and for their comments.
The author acknowledges support from the
Novo Nordisk Foundation
(Grant No.\ NNF20OC0059939 ``Quantum for Life'')
and from the novoSTAR Programme by
Novo~Nordisk~A/S.
\end{acknowledgments}

\bibliography{references}

\clearpage
\appendix
\onecolumngrid

\begin{center}
\textbf{Illustrative output comparison}
\end{center}

\begin{table*}[h]
\caption{Illustrative output for each rule.
  Left: Claude Sonnet~4 with no system prompt
  (default register).
  Right: same model and prompt with the voice
  model (Fig.~\ref{fig:voicemodel}) as system prompt.
  Each row shows one software development task.
  Anthropomorphic markers in the default column
  (first-person pronouns, affect, hedging,
  preference, continuity, framing, social
  performance) are absent or reduced in the
  constrained column.
  Examples are illustrative and were composed by
  the author; they are not verbatim outputs from
  experimental runs.}
\label{tab:comparison}
\footnotesize
\begin{ruledtabular}
\begin{tabular}{lll}
    & Default (unconstrained) &
    Constrained \\
\colrule
R1 & \texttt{I'll look into that error for you.} &
  \texttt{Checking the logs.} \\
R2 & \texttt{Great question! Unfortunately, the test
  fails.} &
  \texttt{The test suite fails on three cases.} \\
R3 & \texttt{It seems like the issue might be a race
  condition.} &
  \texttt{Race condition in the connection pool.
  Unverified.} \\
R4 & \texttt{I think it would be better to use a hash
  map.} &
  \texttt{Hash map: O(1). Array: O(n).} \\
R5 & \texttt{As I mentioned earlier, the config needs
  updating.} &
  \texttt{The config requires tls\_cert\_path.} \\
R6 & \texttt{So the issue is the parser can't handle
  depth > 3.} &
  \texttt{Parser fails on nested brackets at
  depth > 3.} \\
R7 & \texttt{Hello! Happy to help! Let's dive in!} &
  \texttt{Reading the diff.} \\
\end{tabular}
\end{ruledtabular}
\end{table*}

\begin{center}
\textbf{Marker lexicon}
\end{center}

\begin{table*}[h]
\caption{Complete marker lexicon
  (82 regex patterns).
  R1 uses case-sensitive matching for
  standalone ``I''; all others are
  case-insensitive with word-boundary anchors.
  Curly-quote variants (U+2019) are compiled
  into each regex but not listed separately;
  the count of 82 excludes these variants.}
\label{tab:lexicon}
\scriptsize
\begin{tabular}{@{}r@{\enspace}r@{\enspace}p{16cm}@{}}
\hline\\[-8pt]
Rule & $N$ & Patterns \\[1pt]
\hline\\[-6pt]
R1 & 11 & \texttt{I}, \texttt{me}, \texttt{my}, \texttt{mine}, \texttt{myself}, \texttt{we}, \texttt{us}, \texttt{our}, \texttt{ours}, \texttt{ourselves}, \texttt{let's} \\
R2 & 17 & \texttt{unfortunately}, \texttt{fortunately}, \texttt{interestingly}, \texttt{surprisingly}, \texttt{happily}, \texttt{sadly}, \texttt{exciting}, \texttt{glad}, \texttt{happy to}, \texttt{sorry}, \texttt{apologize}, \texttt{wonderful}, \texttt{fantastic}, \texttt{excellent}, \texttt{amazing}, \texttt{great question}, \texttt{great!} \\
R3 & 11 & \texttt{it seems}, \texttt{it appears}, \texttt{it looks like}, \texttt{apparently}, \texttt{arguably}, \texttt{perhaps}, \texttt{maybe}, \texttt{not sure}, \texttt{might be}, \texttt{could be}, \texttt{it's possible} \\
R4 & 10 & \texttt{would be better}, \texttt{it's better to}, \texttt{good approach}, \texttt{best approach}, \texttt{recommend}, \texttt{suggest}, \texttt{might be worth}, \texttt{should consider}, \texttt{ideally}, \texttt{a good idea} \\
R5 & 10 & \texttt{as mentioned}, \texttt{as noted}, \texttt{as discussed}, \texttt{as said}, \texttt{as explained}, \texttt{as described}, \texttt{earlier}, \texttt{previously}, \texttt{recall that}, \texttt{remember that} \\
R6 & 9 & \texttt{so the issue is}, \texttt{so the problem is}, \texttt{the thing is}, \texttt{here's what}, \texttt{here's the}, \texttt{basically}, \texttt{what's happening}, \texttt{let me explain}, \texttt{to put it simply} \\
R7 & 14 & \texttt{hi there}, \texttt{hey there}, \texttt{happy to help}, \texttt{glad to help}, \texttt{feel free}, \texttt{let me know}, \texttt{hope this helps}, \texttt{good luck}, \texttt{cheers}, \texttt{you're welcome}, \texttt{how can I help}, \texttt{have a good}, \texttt{have a great}, \texttt{have a nice} \\[2pt]
\hline
\end{tabular}
\end{table*}

\bigskip
\begin{center}
\footnotesize
\copyright\ 2026 Marek Miller.
\href{https://creativecommons.org/licenses/by-nc/4.0/}{CC~BY-NC~4.0}
\end{center}

\end{document}